\documentclass[twocolumn,showpacs,amsmath,amstex,amssymb,mathfonts,prl,superscriptaddress]{revtex4}
\usepackage{amsthm,amsfonts,graphicx,verbatim}
\usepackage{amsmath}
\usepackage{amssymb}
\usepackage{amsthm}
\usepackage{amsfonts}
\usepackage{listings}
\usepackage{enumerate}
\usepackage{latexsym}
\usepackage{psfrag}
\usepackage{bm}
\usepackage{graphicx}
\usepackage{subfigure}
\newcommand{\be}{\begin{equation}}
\newcommand{\ee}{\end{equation}}
\newcommand{\bea}{\begin{eqnarray}}
\newcommand{\eea}{\end{eqnarray}}

\newcommand{\la}{\langle}
\newcommand{\ra}{\rangle}

\renewcommand{\phi}{\varphi}
\renewcommand{\epsilon}{\varepsilon}

\begin{document}

\title{Measuring entanglement entropy of a generic many-body system with a quantum switch}

\author{Dmitry A. Abanin}
\affiliation{Department of Physics, Harvard University, Cambridge, Massachusetts 02138, USA}

\author{Eugene Demler}
\affiliation{Department of Physics, Harvard University, Cambridge, Massachusetts 02138, USA}


\date{\today}

\begin{abstract}

Entanglement entropy has become an important theoretical concept in condensed matter physics, because it provides a unique tool for  characterizing quantum mechanical many-body phases and new kinds of quantum order. However, the experimental measurement of entanglement entropy in a many-body systems is widely believed to be unfeasible, owing to the nonlocal character of this quantity. Here, we propose a general method to measure the entanglement entropy. The method is based on a quantum switch (a two-level system) coupled to a composite system consisting of several copies of the original many-body system. The state of the switch controls how different parts of the composite system connect to each other. 
We show that, by studying the dynamics of the quantum switch only, the Renyi entanglement entropy of the many-body system can be extracted. We propose a possible design of the quantum switch, which can be realized in cold atomic systems. Our work provides a route towards testing the scaling of entanglement in critical systems, as well as a method for a direct experimental detection of topological order.

\end{abstract}
\maketitle

%
%
%
%
%
%

{\it Introduction.} The concept of entanglement plays a central role in quantum physics and in quantum information science~\cite{quantum_information_review}. While previously entanglement was mostly studied in weakly interacting systems of qubits and photons, more recently it was realized that entanglement is a fundamental property of many-body phases of strongly interacting particles. Mathematically, the degree of entanglement in a pure many-body quantum state is quantified by the {\it entanglement entropy} (EE) defined for a  sub-system ${\cal A}$: the reduced density matrix $\rho_{\cal A}$, obtained by tracing out all degrees of freedom ouside ${\cal A}$, represents some mixed state of ${\cal A}$, and is generally characterized by a non-zero entropy. 

The entanglement entropy provides a valuable tool for characterizing the properties of many-body states. For critical systems, EE shows universal scaling with the sub-system size $l$, with a pre-factor determined by the central charge of the corresponding conformal theory~\cite{Wilczek,CalabreseCardy,Vidal03}. EE also serves as a diagnostic for characterizing new type of quantum order, topological order~\cite{LevinWen, KitaevPreskill}, which cannot be described by the conventional Landau-Ginzburg order parameter. The latter proved particularly useful in recent numerical studies~\cite{Melko,Vishwanath}. 



\begin{figure}
\includegraphics[width=3.3in]{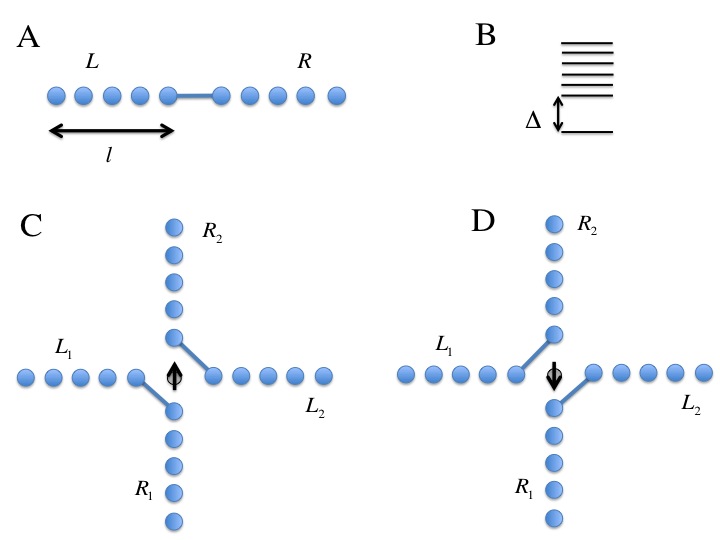}
\vspace{-1mm}
\caption[]{A. The chain represents a general interacting 1D system with short-ranged hopping and interactions. 
We describe a method to measure EE between left (${\cal L}$) and right (${\cal R}$) sub-systems. B. Energy spectrum of the finite chain. The ground state is separated from other excitations by an energy gap $\Delta$, which is given by the thermodynamic gap for gapped  phases, and by the finite size gap for gapless phases, Eq.(\ref{eq:gap_finite}). C, D. Proposed setup for measuring $n=2$ Renyi EE. Two pairs of half-chains ${\cal L}_i$ (${\cal R}_i$) identical to ${\cal L}$ (${\cal R}$) half-chain in A, are arranged in a cross geometry. The quantum switch, positioned at the center of the cross, controls the way in which the two pairs of half-chains are connected by selectively forbidding tunneling to one of the neighbors. 
The overlap of the ground states of the two configurations corresponding to different connections, is directly proportional to $n=2$ Renyi EE, and can be measured by studying Rabi oscillations of the quantum switch. }
\label{fig1}
\end{figure}

Despite the fact that EE has emerged as an indispensable theoretical and numerical tool, it is widely believed that it is nearly impossible to measure it experimentally. This is because EE is a fundamentally non-local quantity; measuring it, seemingly, requires knowing a full reduced density matrix, the size of which grows exponentially with the sub-system size ${\cal A}$. The existing proposals~\cite{KlichLevitov,LeHur}  to measure EE in many-body systems are limited to the special case of non-interacting particles, for which special relations between the reduced density matrix of a sub-system and quantum noise exist; these relations, however, break down when interactions are introduced~\cite{Fradkin}. 

Here we propose a method to measure Renyi entanglement entropies in a general interacting system. We consider a finite 1D chain with short-range hopping and interactions, focusing on the ground state EE  for the system's partition into the left (${\cal L}$) and right (${\cal R}$) parts (Fig.~\ref{fig1}A). The key idea is to engineer a composite system ${\cal L}_i,{\cal R}_i$, $i=1,..,n$, consisting of $n$ copies of the original many-body system, and to couple it to a {\it quantum switch} (a two-level system) in a way described below. By studying the coupled dynamics of the quantum switch and the many-body system, it is possible to extract the EE. Crucially, it is sufficient to measure {\it only the the population of the two states of the quantum switch}. Thus, in principle extremely complicated problem of measuring EE is reduced to studying the dynamics of a single qubit~\footnote{We note that, in spirit, the idea of the quantum switch is somewhat similar to the single atom transistor proposed to study quantum transport and evolution of 1D systems~\cite{Zoller}.}. 

Our proposal is inspired by the works by Horodecky and Ekert~\cite{Ekert02}, as well as by Cardy~\cite{Cardy}. Ref.~\cite{Ekert02} addressed the problem of measuring entanglement in a (mixed) state of several coupled qubits; the method proposed there involved joint operations on different sub-systems, and its complexity grew with the system size. 
Cardy showed that EE in conformally invariant systems can be related to the distribution of energy fluctuations following a quantum quench~\cite{Cardy}. However, experimentally measuring energy fluctuations is a challenging problem, especially in a many-body system, where both kinetic and potential energy of all particles have to be extracted; difficulty of such a measurement also grows rapidly with the system size.

{\it Relation between entanglement entropy and overlaps in a composite system.} We will focus on the $n$th Renyi entropy, defined as follows,
\be\label{eq:Renyi_EE}
S_{n}=\frac{1}{1-n} \log {\rm Tr} \left( \hat\rho_{\cal L}^n \right), 
\ee
where $\rho_{\cal L}$ is the reduced density matrix of the left sub-system. Generally, knowing the Renyi entropies allows one to reconstruct the von Neumann EE by the analytic continuation to $n\to 1$, and to obtain the full entanglement spectrum (the spectrum of the reduced density matrix), see e.g. Ref.~\cite{LeHur}.  

We start with the simplest non-trivial case $n=2$. In what follows, we will rely on the following fact: the Renyi entropy can be related to the overlap of two ground states $|0\ra$, $|0'\ra$ of a composite system that consists of two copies of the original many-body system (such that in total there are four half-chains ${\cal L}_i, {\cal R}_i$, $i=1,2$)~\cite{EE_overlap}. The two configurations correspond to connecting half-chains differently: (i) ${\cal L}_i$ is connected with ${\cal R}_i$, $i=1,2$;  (ii) ${\cal L}_1$ is connected with ${\cal R}_2$, and ${\cal L}_2$ is connected with ${\cal R}_1$, such that the half-chains are swapped. As we shall see below, by coupling quantum switch to the four half-chains, it is possible to extract the overlap $\la 0|0'\ra$ (and therefore EE) from the switch dynamics. 

The origin of this relation is understood by using the Schmidt decomposition of the ground state for a single chain: 
\be\label{eq:schmidt}
|\Psi_0\ra=\sum_i \lambda_i |\psi_i \ra_{\cal L} \otimes  | \phi_i \ra_{\cal R},  
\ee
where $|\psi_i\ra _{\cal L}$, $|\phi_i\ra _{\cal R}$ are the orthogonal wave functions describing left and right sub-systems. 

The ground states  of the composite system are given by the tensor product of two chains' ground states,
\be\label{eq:GS00'}
|0\ra=|\Psi_0\ra _{{\cal L}_1,{\cal R}_1} \otimes  |\Psi_0\ra _{{\cal L}_2,{\cal R}_2}, \,\, \,\,  |0'\ra=|\Psi_0\ra _{{\cal L}_1,{\cal R}_2} \otimes  |\Psi_0\ra _{{\cal L}_2,{\cal R}_1}.
\ee
They appear similar, the only difference being that in the $|0\ra$ state ${\cal L}_1$, ${\cal R}_1$ and ${\cal L}_2$, ${\cal R}_2$ pairs are entangled, while in the $|0'\ra$ state we need to swap the right sub-systems. Applying the Schmidt decomposition (\ref{eq:schmidt}) for each $|\Psi_0\ra$ in the above equation, we obtain
\be\label{eq:0}
|0\ra=\left( \sum_i \lambda_i |\psi_i \ra_{{\cal L}_1} \otimes  | \phi_i \ra_{{\cal R}_1} \right)\otimes  \left( \sum_j \lambda_j |\psi_j \ra_{{\cal L}_2} \otimes  | \phi_j \ra_{{\cal R}_2} \right). 
\ee
\be\label{eq:0'}
|0'\ra=\left( \sum_i \lambda_i |\psi_i \ra_{{\cal L}_1} \otimes  | \phi_i \ra_{{\cal R}_2} \right)\otimes  \left( \sum_j \lambda_j |\psi_j \ra_{{\cal L}_2} \otimes  | \phi_j \ra_{{\cal R}_1} \right).
\ee
The overlap of the two ground states, which can be evaluated by using this representation, is given by:
\be\label{eq:overlap}
\la 0| 0'\ra=\sum_i \lambda_i ^4={\rm Tr} \left( \hat\rho_{\cal L}^2 \right)= e^{-S_2},  
\ee
and therefore is related to the $S_2$ EE of a single many-body system. 

{\it Proposed setup.} We consider the following realization of the composite system: two left and two right sub-systems, arranged in a cross geometry, as illustrated in Fig.~\ref{fig1}C. Initially, each ${\cal L}_i$, $i=1,2$ sub-system is connected to both right sub-systems ${\cal R}_1, {\cal R}_2$. In the center of the cross, a quantum switch is placed -- a two-level system with states $|\uparrow\ra$ and $|\downarrow\ra$, which controls the connection between different sub-systems (see Fig.~\ref{fig1}C,D). When the switch is in the $|\uparrow\ra$ state, tunneling between ${\cal L}_1,{\cal R}_2$ pair and ${\cal L}_2,{\cal R}_1$ pair is blocked, such that configuration (i) is realized; when the switch is in the $|\downarrow\ra$ state, tunneling between ${\cal L}_1,{\cal R}_1$ pair and ${\cal L}_2,{\cal R}_2$  is blocked, corresponding to the configuration (ii). 

\begin{figure}
\includegraphics[width=3.3in]{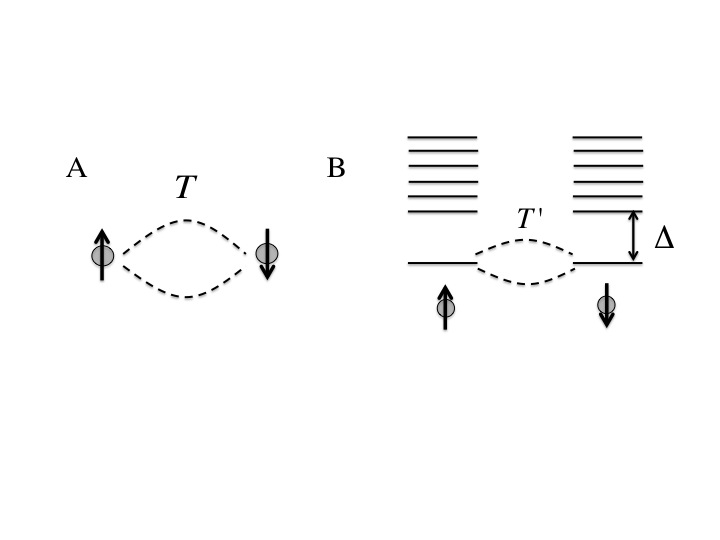}
\vspace{-15mm}
\caption[]{The dynamics of the coupled switch-chain system can be used to measure $n=2$ Renyi entropy. A. We introduce dynamics of the quantum switch by turning on weak tunneling with amplitude $T$ between the $|\uparrow\ra$ and $|\downarrow\ra$ states (A). B. Schematic of the energy spectrum of the switch-chains system. The ground state is doubly degenerate and is separated from the excited states by a gap. Assuming that $T\ll \Delta$, the system will perform Rabi oscillations between the two ground states, with the renormalized effective tunneling between them, see Eq.(\ref{eq:low_energy}). By measuring the renormalized Rabi frequency, it is possible to extract the overlap of the ground states of the two configurations of the many-body system shown in Fig.~\ref{fig1}C-D, and therefore the Renyi entropy. }
\label{fig1add}
\end{figure}

First, we assume that the two states of the quantum switch are completely decoupled (later on, we will introduce the dynamics). In this case, the spectrum of the switch-chains system consists of two sectors, corresponding to $|\uparrow\ra$, $|\downarrow\ra$ states of the switch, as illustrated in Fig.~\ref{fig1add}B. The spectrum in each sector can be related to the spectrum of the single many-body system $\{ E_i, |\Psi_i \ra \}$ ($E_i$ being eigenenergy, $|\Psi_i\ra$ the corresponding wave function): the eigenfunctions are given by the tensor products of the eigenfunctions for a single chain. For configuration (i), the spectrum has the following form,
\be\label{eq:spectrum_i}
\left\{E_i+E_j ,  \,\, |\uparrow\ra \otimes |\Psi_i\ra_{{\cal L}_1, {\cal R}_1} \otimes  |\Psi_j\ra_{{\cal L}_2, {\cal R}_2} \right\}, \,\, i,j=0,1,2,...
\ee
while for configuration (ii) it is given by
\be\label{eq:spectrum_ii}
\left\{E_i+E_j, \,\,  |\uparrow\ra \otimes |\Psi_i\ra_{{\cal L}_1, {\cal R}_2} \otimes  |\Psi_j\ra_{{\cal L}_2, {\cal R}_1} \right\}, \,\, i,j=0,1,2,...
\ee

 \begin{figure}
\includegraphics[width=3.3in]{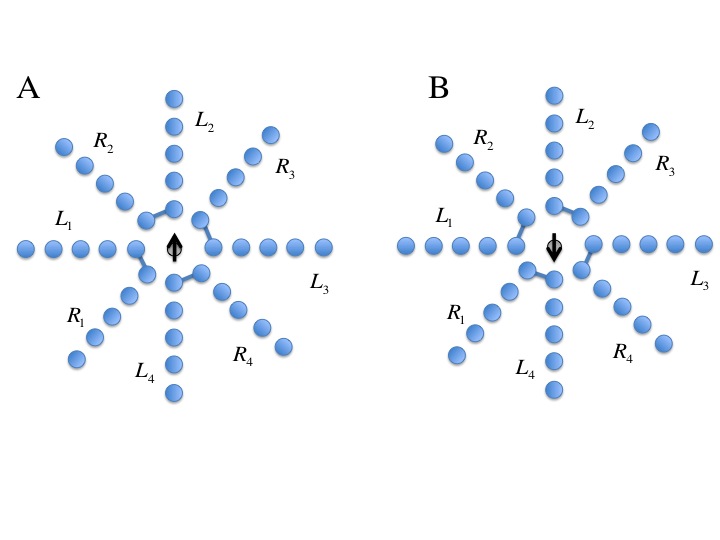}
\vspace{-1mm}
\caption[]{Setup proposed for measuring higher Renyi EE. Case $n=4$ is shown. $2n$ half-chains ${\cal L}_i$, ${\cal R}_i$ are arranged in a star geometry in an alternating fashion. Initially, tunneling to both neighboring half-chains is allowed. The quantum switch blocks tunneling to one of the neighbors; thus, the two states of the switch realize two configurations, in which the half-chains are connected differently (each ${\cal L}_i$ is connected to its clockwise neighbor ${\cal R}_i$ for $|\uparrow\ra$ state, and it is connected to the counter-clockwise neighbor ${\cal R}_{i+1}$ for $|\downarrow\ra$ state.)}
\label{fig2}
\end{figure}

The ground state is doubly degenerate and is separated from the rest of the spectrum by a gap. The gap $\Delta$ is given by the thermodynamic gap for gapped quantum phases (up to small finite-size corrections). For a gapless phase, the gap is due to the finite size of the chain, and is given by 
\be\label{eq:gap_finite}
\Delta\sim \frac{\hbar v}{l}, 
\ee
where $v$ is the velocity of gapless excitations, and $l$ is the size of the chain. The wave functions of the two ground states are given by:
$$|GS\ra=|\uparrow\ra \otimes |0\ra, \,\,\, |GS'\ra=|\downarrow\ra \otimes |0'\ra$$.

In order to extract the overlap $\la 0|0'\ra$, we now introduce weak tunneling between the two states of the quantum switch, adding the following term in the Hamiltonian:
\be\label{eq:tunneling}
H_t=T ( |\uparrow\ra \la \downarrow|+|\downarrow\ra \la \uparrow|). 
\ee
Such tunneling gives rise to the hybridization of the two ground states $|GS\ra=|\uparrow\ra\otimes |0\ra$, $|GS'\ra=|\downarrow\ra\otimes |0'\ra$. Assuming small tunneling amplitude $T\ll \Delta$, we can only consider two lowest states, and the effective low-energy Hamiltonian describing their dynamics takes the following form: 
\be\label{eq:low_energy}
H_{eff}=\tilde{T}  ( |GS\ra \la GS'|+|GS'\ra \la GS|), \,\, \tilde{T}=T \la 0 |0' \ra. 
\ee
Thus, the renormalization of the tunneling amplitude is proportional to the desired overlap. 

\begin{figure}
\includegraphics[width=3.4in]{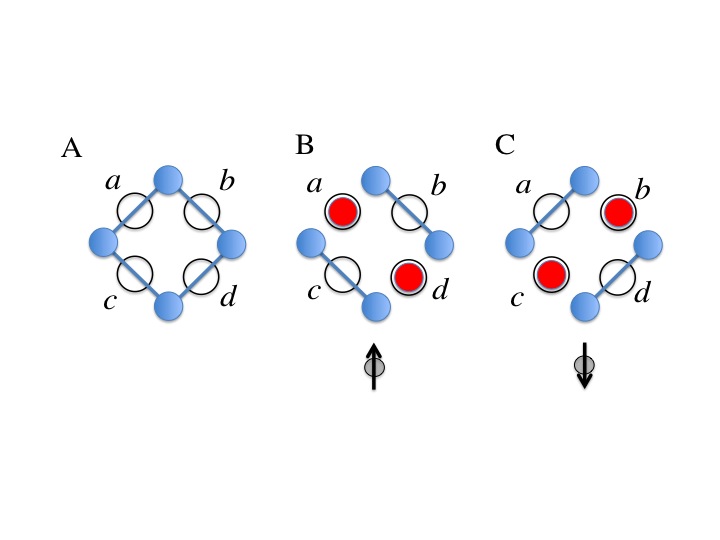}
\vspace{-15mm}
\caption[]{
Proposed design of the quantum switch in cold atomic systems. Between the end sites of the four half-chains four quantum wells $a,b,c,d$ are positioned. When empty, they do not affect the tunneling between the end sites (A). Two dipolar molecules (red circles) interacting via long-range repulsive interactions with each other are put in the quantum wells (B,C). The ground state of the molecules is doubly degenerate, corresponding to the occupation of either $a,c$ wells (B), or $b,d$ wells (C), such that the repulsive interactions between them are minimized. The dipolar molecules are assumed to be interacting with the particles constituting the many-body chains, such that depending on their state they can block the tunneling between neighboring sites (B,C). The two states represent the $|\uparrow\ra$ and $|\downarrow\ra$ states of the quantum switch.}
\label{fig3}
\end{figure}

The renormalization of the tunneling amplitude can be experimentally measured by studying the Rabi oscillations of the switch. One possible way to measure the Rabi frequency is as follows: initially, $T=0$ (two switch levels are decoupled), and the system is prepared in the first ground state $|GS\ra$. At time $t=0$, the tunneling is switched on. The system will oscillate between the two states $|GS\ra, |GS'\ra$, such that the difference of probabilities of the two states is given by:
\be\label{eq:population_imb}
P_{GS}-P_{GS'}=\cos (\tilde\Omega t), \,\, \tilde\Omega=\tilde{T}/\hbar .
\ee 
By measuring the population of the two states of the quantum switch, the renormalized Rabi frequency can be extracted. This gives the desired overlap, and therefore the Renyi entropy via Eq.(\ref{eq:overlap}). 

{\it Measuring higher Renyi entropies $S_n$, $n>2$}.  The method described above can be extended to measure the higher Renyi entropies. The setup is as follows: $n$ left and $n$ right half-chains ${\cal L}_i, {\cal R}_i$ are arranged in a star geometry in an alternating fashion, such that the ${\cal L}_i$ half-chain neighbors ${\cal R}_i$ and ${\cal R}_{i+1}$ half-chains (as illustrated in Fig.~\ref{fig2} for case $n=4$). Initially, each half-chain is coupled to both of its neighbors. The quantum switch can selectively block the tunneling to one of the neighbors; thus, similarly to the case $n=2$, the two states of the quantum switch realize different configurations of the composite system: in the first configuration, ${\cal L}_i$ is connected to ${\cal R}_i$, while in the second, ${\cal L}_i$ is connected to ${\cal R}_{i+1}$ for $1\leq i \leq n-1$, and ${\cal L}_n$ is connected to ${\cal R}_1$. 

The overlap of the ground state wave functions $|0_n\ra$, $|0'_n\ra$ of the two configurations is directly related to the $n$th Renyi entropy (see, e.g., Ref.~\cite{Cardy}): 
\be\label{eq;Sn}
S_n=\frac{1}{1-n}\log \la 0_n | 0'_n\ra. 
\ee
The overlap and $S_n$ can be measured in a Rabi experiment, as for the case $n=2$. 

{\it A possible design of the quantum switch.} Although the proposed setup is generic and one can envision its realization in a number of solid state and atomic physics systems, we believe that it could be most easily implemented in systems of cold atoms. The 1D chains described by the transverse field Ising model have been recently realized in such systems~\cite{Greiner}. Below we propose one possible design of a quantum switch in cold atomic systems. It can be used, for example, to test the universal scaling of EE with the system size in the transverse field Ising model.

The design of quantum switch, illustrated in Fig.~\ref{fig3}, involves two dipolar molecules which interact repulsively with each other and with the particles that constitute the many-body system. The dipolar molecules reside in a four-well potential arranged in a square pattern, with the vertices of the square situated on the lines connecting last sites of the neighboring half-chains (see Fig.~\ref{fig3}). 
Neglecting tunneling between the wells, the ground state of the molecules is doubly degenerate, and corresponds to the particles occupying the opposite vertices of the square. 

We impose two main requirements: first, the dipolar molecules interact strongly with the particles in the many-body system, such that when a given quantum well is occupied, the tunneling between the half-chains that neighbor that well is blocked. Second, the interactions between the dipolar molecules must be strong enough such that we can neglect the excited states of the molecules (e.g., a configuration in which they occupy neighboring quantum wells). Under these demanding, but realistic conditions, the four-well system provides a version of a quantum switch. Experimentally, one would measure the dynamics of such a switch by monitoring the occupation of different quantum wells as a function of time. 

This switch design can be extended for the case $n>2$. $2n$ wells should be designed between the ends of the half-chains shown in Fig.~\ref{fig2}. They should be populated by $n$ dipolar molecules with such long-range interactions that in the ground state every other well is occupied. Then, the two degenerate ground states of the dipolar molecules correspond to the $|\uparrow\ra$, $|\downarrow\ra$ states of the quantum switch. 

{\it Conclusions and generalizations.} In conclusion, we have proposed a method to experimentally measure the ground state EE in a generic many-body system. We have considered a specific setup which can be realized with current experimental means in cold atomic systems. We expect that the approach proposed here will enable the tests of the universal scaling of EE in various critical many-body systems. 

The ideas presented above can also be generalized to measure entanglement entropy in two-dimensional systems~\cite{SOM}. Such a measurement provides a direct experimental test of topological order~\cite{LevinWen,KitaevPreskill} characteristic of systems such as spin liquids and fractional quantum Hall states. 

Finally, we note that the proposed approach can be used to measure EE of an arbitrary excited states, as long as they are separated from the other states by an energy gap. Exploring the EE of excited states may shed light on the nature of localization in disordered many-body systems. 

 
 {\it Acknowledgements.} We thank M. Greiner, M. Lukin, P. Zoller, A. Vishwanath, and A. Yacoby for insightful discussions, and acknowledge support from
Harvard-MIT CUA, NSF Grant No. DMR-07-05472, DARPA OLE
program, AFOSR Quantum Simulation MURI, and the ARO-MURI on Atomtronics.

\bibliography{paper}

\begin{figure}
\vspace{-1cm}
\includegraphics[width=2.5in]{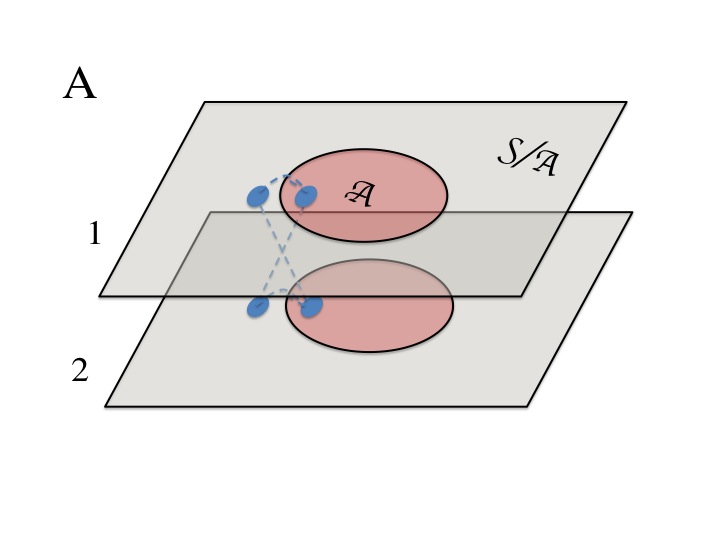}
\vspace{-1cm}

\includegraphics[width=2.5in]{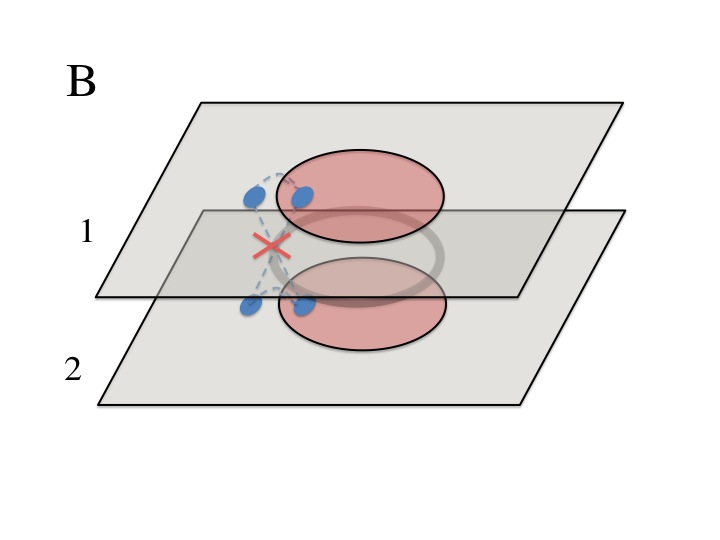}
\includegraphics[width=2.5in]{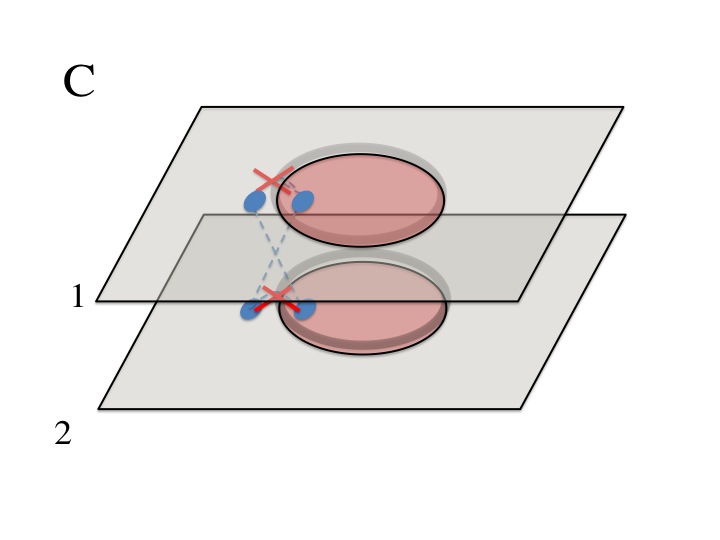}
\vspace{-1cm}
\caption[]{Setup for measuring $S_2$ for a finite sub-system ${\cal A}$ of a two-dimensional system.   A. Two copies of the system are engineered, and are displaced by some distance in the $z$ direction. The lattice potential is engineered in such a way that tunneling can occur between ${\cal S\backslash A}$ and ${\cal A}$ both within systems 1,2, and between systems 1 and 2. B. Quantum switch in state $|\uparrow\ra$ quenches the tunneling between 1 and 2, such that the two systems are decoupled. C. Quantum switch in state $|\downarrow\ra$ blocks the tunneling between ${\cal S\backslash A}$ and ${\cal A}$ within 1 and 2, such that ${\cal S\backslash A}$ in 1 connects to ${\cal A}$ in 2, and vice versa. The Rabi oscillations of the quantum switch can then be used to measure $S_2$ for region ${\cal A}$, in close analogy to the 1d case. }
\label{figS}
\end{figure}

\section{Supplementary Online Material.}

Here we discuss a setup for measuring entanglement entropy of 2d systems. One important application of such a measurement would be to detect topological order and its type in gapped systems~\cite{LevinWen,KitaevPreskill} We consider two copies, 1 and 2, of a 2d quantum system, positioned above each other, as illustrated in Fig.~\ref{figS}; we focus on measuring $n=2$ Renyi entropy of some finite sub-system ${\cal A}$. 

As for one-dimensional case, we assume that the Hamiltonian of the quantum system is such that the hopping is short-ranged. 
The composite  system is engineered in such a way that the tunneling across the boundary of region ${\cal A}$ can take place either within system 1 and within system 2, or it can take the particles between systems 1 and 2 (Fig.~\ref{figS}A). Apart from the regions near the boundary, 1 and 2 are decoupled. To engineer such a system is a challenging, but in principle solvable problem. 

Another crucial ingredient is the quantum switch, which can selectively block the tunneling across the boundary. In the $|\uparrow\ra$ state, the switch blocks the tunneling between 1 and 2, such that the two copies become completely decoupled (Fig.~\ref{figS}B). 
In the $|\downarrow\ra$ state, the switch blocks tunneling across the boundary within system 1 and within system 2. Therefore, the systems are "swapped" -- $\cal{S\backslash A}$ in system 1 now connects with ${\cal A}$ in system 2 and vice versa. 
In analogy to 1d case, the Rabi oscillations of the quantum switch can be used to measure the Renyi entanglement entropy. 

One possible design of the quantum switch in the 2d case is to have two dipolar molecules, which are trapped in a wave guide that has a shape of the boundary of region ${\cal A}$ in the $x-y$ plane. In the $z$ direction, the trapping potential is such that the molecules have two ground states: the one in which the wave function is localized between the planes in which systems 1 and 2 are situated, and the one where it can form a superposition of states localized in the planes of 1 and 2 (see Fig.{\ref{figS}B,C}). Making such configurations degenerate requires a special choice of the potential profile for dipolar molecules, as well as tuning the interactions between them. Detailed design of such quantum switch is beyond the scope of this paper; here our goal is to mostly illustrate that such measurement is possible n principle.

\end{document}